# Learning Game Representations from Data Using Rationality Constraints

**Xi Alice Gao**
School of Engineering and Applied Sciences
Harvard University
Cambridge, MA 02138
xagao@seas.harvard.edu

**Avi Pfeffer**
Charles River Analytics Inc.
apfeffer@cra.com

## Abstract

While game theory is widely used to model strategic interactions, a natural question is where do the game representations come from? One answer is to learn the representations from data. If one wants to learn both the payoffs and the players' strategies, a naive approach is to learn them both directly from the data. This approach ignores the fact the players might be playing reasonably good strategies, so there is a connection between the strategies and the data. The main contribution of this paper is to make this connection while learning. We formulate the learning problem as a weighted constraint satisfaction problem, including constraints both for the fit of the payoffs and strategies to the data and the fit of the strategies to the payoffs. We use quantal response equilibrium as our notion of rationality for quantifying the latter fit. Our results show that incorporating rationality constraints can improve learning when the amount of data is limited.

## 1 Introduction

Game theory has been widely used to model strategic interactions of agents. A natural question is where do the game representations come from? Traditionally, games have usually been constructed by hand, but in that case they may only be theoretical abstractions of reality. Alternatively, one might have access to a simulator, for example, in the trading agent competition, but that is unusual. The problem is similar to that of developing a probabilistic model, and a common approach, which we take in this paper, is to learn the model from data.

Consider for example the following situation. A farmer wants to begin selling fruits in a market. She has to decide where in the market to locate her store, which types of fruits to sell, and what prices to charge for the fruits. She does not know what the other farmers currently do nor what payoffs the other farmers are receiving for their decisions. This is a game and our new farmer would like to learn both the rules of the game and the strategies that the other players play from observing them. Moreover, she would like to learn these quickly with limited data.

A naive approach to this learning problem is to use the maximum likelihood estimation to learn the strategy profiles and payoff values separately from observed data. However, this approach is prone to over-fitting since limited amount of data may not accurately reflect the actual strategies and payoffs. More importantly, this approach ignores the fact that agents might already be playing reasonably good strategies. It does not connect the strategies played by the agents to their payoffs while learning. Our main contribution is to make this connection, as illustrated in Figure 1.

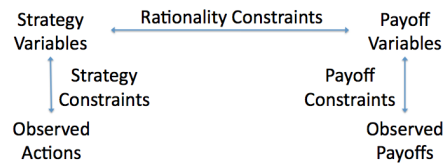

Figure 1: Our Learning Approach

The naive approach makes the connection between the strategy and payoff variables with the observed data, but does not connect the strategy variables to the payoff variables. An alternative approach could learn the payoffs from data, and then connect the strategies to the payoffs by solving the game, using some notion of rationality. This approach, however, ignores the observed data about strategies. Our approach makes all three connections. We learn both the payoff variables and strategy variables from data, and use the payoffs to inform the strategies and vice versa.

A possible notion of rationality is Nash equilibrium. However, Nash equilibrium is a brittle concept since it assumes players have completely converged to playing rational strategies and does not allow for noise. Instead, we use the quantal response equilibrium concept to model the rationality of agents. In quantal response equilibrium, players do not choose best responses with probability one as in Nash equilibrium. Instead, they "better respond" by choosing responses with higher expected payoffs with higher probabilities. Using the quantal response equilibrium allows us to model bounded rationality of agents as well as being robust to noise in the data.

Our approach is formulated as a weighted constraint satisfaction problem (WCSP). Our WCSP model has variables for both mixed strategy profiles and payoff values. The two sets of soft constraints assign lower costs to variable assignments that have higher probability based on the observed data. The third set of soft constraints expresses the bounded rationality of the agents using the quantal response equilibrium concept. The advantage of using the WCSP framework is that we are able to incorporate and balance the goals of fitting the learned strategy profiles and payoff values to the observed data with the goal of connecting the learned strategy profiles and payoff values using the quantal response equilibrium. Furthermore, using a CSP framework allows us to potentially add additional types of constraints for dealing with more complex game structures. For instance, one possible future work is to add additional constraints for learning the structures of graphical games (Kearns et al., 2001).

**Related Work.** There has been some research in recent years on learning games from data. Ficici et al. (2008) learned a reduced game form representation from data by clustering agents with similar strategic views of the game and used this representation to approximately solve asymmetric games of many players. Vorobeychik et al. (2007) used regression learning techniques to learn payoff functions of infinite games. Duong et al. (2009) applied existing learning algorithms such as branch-and-bound, greedy, and local search algorithms to learning graphical games from data. These search algorithms were evaluated based on metrics such as minimizing empirical loss, approximating graphical structure, and approximating Nash equilibria. All these approaches differ from ours in that they do not use rationality to inform the learning.

Soni et al. (2007) formulated the problem of computing pure strategy approximate Nash equilibria in one-shot complete-information games as a constraint satisfaction problem. Their PureProp CSP defined variables corresponding to the strategy profiles of players and constraints to solve for $\epsilon$-Nash equilibrium of the resulting game. Their work is similar to our approach since we essentially use rationality constraints to solving for quantal response equilibrium of the normal form game based on observed data on payoff values. However, their work differs from our approach because we also use constraints related to learning the parameters from observed data in our CSP framework.

## 2 Preliminaries

### 2.1 Normal Form Games

In this paper, we consider one-shot complete information normal form games. Such a game $\mathbf{G} = [\mathbf{I}, \mathbf{A}, \mathbf{U}]$ is given by a set of players $\mathbf{I}$ with $|\mathbf{I}| = N$, a set of pure strategies $\mathbf{A} = \times_{i \in \mathbf{I}} A_i$ where $A_i$ is the set of pure strategies for player $i \in \mathbf{I}$, and $\mathbf{U} = \times_{i \in \mathbf{I}} u_i$ where $u_i : \mathbf{A} \to \Re$ is the payoff function for player $i \in \mathbf{I}$. A mixed strategy of player $i$ is denoted by $\sigma_i : A_i \to [0, 1]$ where $\sum_{a \in A_i} \sigma_i(a) = 1$ is a probability distribution over all the pure strategies of player $i$. We use $-i$ to refer to the set of all players other than player $i$. Thus, $a_{-i}$ and $\sigma_{-i}$ denote the joint pure strategy and mixed strategy profile of the set of players $-i$.

We focus on finite games for which the set of players $\mathbf{I}$ and the sets of pure strategies $A_i$ for each player $i \in \mathbf{I}$ are finite. The game $\mathbf{G}$ is considered complete information in the sense that each player $i$ has full knowledge of the payoff functions of the other players.

### 2.2 Equilibrium Notions

In our learning model, we consider two popular equilibrium notions: the mixed strategy Nash equilibrium and the quantal response equilibrium.

Nash equilibrium assumes that every player is completely rational. In Nash equilibrium, every player $i$ is playing a best response to the mixed strategy profile of all the other players $-i$.

**Definition: 1** *A joint mixed strategy profile $\sigma^*$ is Nash equilibrium if every player $i$ is playing a mixed strategy best response to the strategy profiles of its opponents $-i$, i.e.*

$$u_i(\sigma^*_{-i}, \sigma^*_i) \geq u_i(\sigma^*_{-i}, \sigma_i), \forall \sigma_i \quad (1)$$

This definition implies that no player can benefit in expectation by deviating unilaterally from the Nash equilibrium strategy profile $\sigma^*$.

We use the quantal response equilibrium notion to model the bounded rationality of agents (McKelvey and Palfrey, 1996). In a quantal response equilibrium, players fix their strategies, form beliefs about the strategies of the other players, compute their expected payoffs based on their beliefs, and choose a strategy which assigns relatively higher probabilities to actions with relatively higher expected payoffs. We use the most common specification for the quantal response equilibrium called the logit equilibrium (LQRE). The LQRE uses an exponentiated response function and a positive real valued parameter $\lambda$ to capture how sensitive each player is to the differences in expected payoffs. The parameter $\lambda$ can be interpreted as a measure of the degree of rationality of the particular player choosing this response function. For instance, when $\lambda = 0$, the player chooses each pure strategy with equal probability. As $\lambda$ increases, the player becomes more and more responsive to the differences in expected payoffs of different pure strategies and chooses his mixed strategy best response accordingly. The strategies chosen by the players in a LQRE converges to Nash equilibrium when $\lambda$ approaches infinity.

**Definition: 2** *A joint mixed strategy profile $\sigma^*$ is a LQRE for some non-negative real number $\lambda$ (referred to as $\lambda$-LQRE hereafter), if for every player $i$, $\sigma^*(a_i)$ for every $a_i \in A_i$ is a exponentiated best response to the joint mixed strategy profile $\sigma^*_{-i}$, i.e. for all $a_i \in A_i$, we have*

$$\sigma_i^*(a_i) = \frac{e^{(\lambda \sum_{a_{-i} \in A_{-i}} \sigma^*_{-i}(a_{-i}) u_i(a_i, a_{-i}))}}{\sum_{a_j \in A_i} e^{(\lambda \sum_{a_{-i} \in A_{-i}} \sigma^*_{-i}(a_{-i}) u_i(a_j, a_{-i}))}} \quad (2)$$

### 2.3 Weighted Constraint Satisfaction Problems

A classic constraint satisfaction problem (CSP) is defined by a set of variables, each with a finite domain, and a set of constraints. Each constraint consists of a subset of the variables, and a relation over the variables in the constraint. A solution to a CSP is an assignment of values to the variables such that every constraint is satisfied, i.e., the projection of the assignment onto the constraint variables is in the constraint relation.

A weighted CSP (WCSP) is similar to a CSP except that it has two kinds of constraints. Hard constraints are like those in an ordinary CSP. Soft constraints consist of a set of variables and a cost function from those variables to non-negative numbers. The goal of solving a CSP is to find an assignment such that all the hard constraints are satisfied, while the sum of the costs of the soft constraints for the assignment is minimized.

## 3 Problem Definition

We consider generating $M$ samples of the play of a one-shot simultaneous-move game **G**. **G** is a complete information game, so the players have complete knowledge of the payoff matrix of the game. We assume that player $i$ chooses a fixed mixed strategy $\sigma_i$ chosen in advance and plays it for every sample generated. This is a simplifying assumption that allows us to treat the training instances as independent; relaxing it is a topic for future work. For the $k$-th sample, player $i$ chooses to play some pure strategy $a_i \in A_i$ and incurs some real valued payoff $u_i(\mathbf{A})$ where $\mathbf{A}$ is the joint pure strategy profile chosen by all the players. For example, in our fruit vendor example, the players are the fruit vendors, and the pure strategies of each player are the location of the stall, the types of fruit sold, and the prices. This game is played for many times by the fruit vendors.

In this paper, we take the perspective of an outside observer (referred to as $O$ hereafter). $O$ is not a participant of the game but might be someone who potentially wants to join the game, just like the fruit seller who wants to join the market. We assume that the number of players and the set of pure strategies available to each player are known to $O$. We also assume that the set of players playing the game is fixed. Given these assumptions, $O$ observes the game being played for $M$ times. Whenever, the players choose a joint pure strategy profile $\mathbf{a}$ for the $k$-th sample, $O$ can observe this joint pure strategy profile $\mathbf{a}$ perfectly. This assumption fits our domain, where the observer can observe the fruit, prices and location of each vendor, but can easily be relaxed if necessary. However, $O$ cannot observe the payoffs directly, but only has a noisy estimate of them. In the fruit example, the observer can observe the number of customers at each stall but not the exact transactions. We model this by having $O$ observe a payoff that is Gaussian distributed around its expected value $u_i(\mathbf{A})$ using a fixed variance $R^2$; other noise models can easily be accommodated.

Our goal is, given these observations, to learn a representation of the game that balances the fit of the learned game to the data with the assumption that the players are playing a $\lambda$-LQRE for some fixed $\lambda$. This balancing involves three factors: (1) the fit of the estimated payoffs to the observed payoffs; (2) the fit of the estimated strategies to the observed strategies; and (3) the fit of the estimated strategies to the estimated payoffs. The $\lambda$-LQRE assumption may or may not be correct. Therefore, we do not enforce this assumption in a hard manner, but rather introduce slack, penalizing estimates that deviate far from it. The assumption serves as a force with which to enforce some sense on the data. It behaves rather like a Bayesian prior. With a large amount of data, the LQRE assumption will be overcome, but with small amounts of data it will help prevent over-fitting. For the LQRE assumption, we assume that the outside observer knows all the $\lambda$ values for the players although the players do not need to know each other's $\lambda$ values.

## 4 Our WCSP Learning Model

We formulate our learning model using the weighted constraint satisfaction framework. Our WCSP model has one variable corresponding to each $\sigma_i(a_i)$ which is the probability of the pure strategy $a_i$ in the mixed strategy $\sigma_i$ of each player $i$. Moreover, our WCSP also defines one variable for each entry in the $u_i(\mathbf{A})$ which is the true payoff vector of player $i$ when the joint pure strategy profile chosen by all the players is $\mathbf{A}$. Since the domains of the strategy and payoff variables are continuous, we discretize them to be multiples of real valued discretization parameters $\epsilon$ and $\delta$ respectively. We also restrict the payoff values $u_i(\mathbf{A})$ to be within a pre-defined range $[u_{min}, u_{max}]$ where $u_{min}, u_{max} \in \Re$ are the minimum and the maximum of the observed payoff values. Now we define the four types of constraints in our WCSP learning model.

**Strategy Maximum Likelihood Constraints**

Intuitively, the value of $\sigma_i(a_k)$ for mixed strategy $\sigma_i$ for player $i$ and the pure strategy $a_k \in A_i$ for player $i$ gives the probability of observing the pure strategy $a_k$ being played by player $i$ for a particular instance of game play. Therefore, given a sequence of observed pure strategies $a_k^1, ..., a_k^M$ chosen by player $i$, we can calculate the log-likelihood of a particular $\sigma_i(a_k)$ value given the observed sequence of pure strategies as

$$\log\left(\sigma_i(a_k^1) \times ... \times \sigma_i(a_k^M)\right) = \sum_{j=1; a_k^j = a_k}^{M} \log \sigma_i(a_k^j) \quad (3)$$

Given a sequence of observed pure strategies, for each player $i$, we would like to choose a value for $\sigma_i(a_k)$ in order to maximize the log-likelihood of the observations. Maximizing the log-likelihood of observations is equivalent to minimizing the negative log-likelihood of the observations.

Therefore, for our strategy maximum likelihood constraints, we associate a cost with each possible discretized value of $\sigma_i(a_k)$ for each pure strategy $a_k \in A_i$ of player $i$, which is equal to the negative log-likelihood of the observed sequence of pure strategies chosen by player $i$ for the $M$ samples of game play. For every player $i$, and for each pure strategy $a_k \in A_i$ for player $i$, the cost can be calculated as follows:

$$cost(\sigma_i(a_k)) = -\log P(a_i^1, ..., a_i^M | \sigma_i(a_k)) \quad (4)$$

Based on the above equation, the value of $\sigma_i(a_k)$ associated with minimum cost has the highest probability to be the true value of $\sigma_i(a_k)$ in the true mixed strategy $\sigma_i(a_k)$ chosen by player $i$.

**Strategy Consistency Constraints**

This consistency constraint ensures that the values of $\sigma_i(a_k)$ form a valid mixed strategy probability distribution since our strategy maximum likelihood constraints were defined for each pure strategy separately rather than for the set of pure strategies for a particular player.

The strategy consistency constraint is a unary hard constraint for the variables $\sigma_i(a_k)$ of player $i$. This constraint specifies that the probabilities $\sigma_i(a_i)$ assigned to the pure strategies $a_i \in A_i$ must sum to 1, i.e.

$$\forall i \in \mathbf{I}, \sum_{a_k \in A_i} \sigma_i(a_k) = 1 \quad (5)$$

**Payoff Maximum Likelihood Constraints**

We defined the observed payoff of a player as a Gaussian distribution $N(u_i(\mathbf{A}), R^2)$. Therefore, the log-likelihood of observing a sequence of payoff values $v_i^1, ..., v_i^M$ with the corresponding pure strategy profiles $a^1, ..., a^M$ for player $i$ can be calculated as

$$cost(u_i(\mathbf{a})) = \sum_{j=1: \mathbf{a^j} = \mathbf{a}}^{M} \log N(v_i^j, \mathbf{a^j} | u_i(\mathbf{a}), R^2) \quad (6)$$

Given a sequence of observed payoff values, for each player $i$, we would like to choose a value for $u_i(\mathbf{a})$ in order to maximize the log-likelihood of the observations, which is equivalent to minimizing the negative log-likelihood of the observations.

For our payoff maximum likelihood constraints, we define the cost for $u_i(\mathbf{a})$ as the negative log-likelihood of the observed sequence of payoffs of player $i$ when the observed pure strategy profile $\mathbf{a^j}$ is the same as $\mathbf{a}$. For every player $i$, the cost can be calculated as follows:

$$cost(u_i(\mathbf{a})) = -\sum_{j=1: \mathbf{a^j} = \mathbf{a}}^{M} \log N(v_i^j, \mathbf{a^j} | u_i(\mathbf{a}), R^2) \quad (7)$$

**Rationality Constraints**

There is a rationality constraint for every pure strategy of every player. This constraint only involves variables and does not involve any of the observed data. For the rationality constraint, the cost for player $i$ for a pure strategy measures how far off the probability of the strategy is from its quantal response probability. From Equation 2, we obtain that in a $\lambda$-LQRE,

$$\left| \exp(\lambda \sum_{a_{-i}} \sigma_{-i}(a_{-i}) u_i(a_k, a_{-i})) - \sigma_i(a_k) \sum_{a_k \in A_i} \exp(\lambda \sum_{a_{-i}} \sigma_{-i}(a_{-i}) u_i(a_k, a_{-i})) \right| = 0 \quad (8)$$

We therefore obtain the following constraint for all $i \in \mathbf{I}$ and $a_k \in A_i$:

$$cost(\sigma_i(a_k), \sigma_{-i}(\cdot), u_i(a_k, \cdot)) = \quad (9)$$
$$\alpha | \exp(\lambda \sum_{a_{-i}} \sigma_{-i}(a_{-i}) u_i(a_k, a_{-i}))$$
$$-\sigma_i(a_k) \sum_{a_k \in A_i} \exp(\lambda \sum_{a_{-i}} \sigma_{-i}(a_{-i}) u_i(a_k, a_{-i})) |$$

The $\sigma_{-i}(\cdot)$ indicates that the constraint has a variable for every pure strategy of every other player. Similarly, $u_i(a_k, \cdot)$ indicates that it has one for every action profile consistent with player $i$ playing $a_k$. Clearly, this is a lot of variables, so the constraint will be huge. We will discuss how to deal with this in the next section. Note that our LQRE constraints only include strategy profiles observed in the data.

The parameter $\alpha$ determines the strength of the rationality constraints compared to the data-driven strategy and payoff constraints. A larger $\alpha$ is analogous to having a stronger prior, and will make it more difficult to overcome the rationality constraint, whereas a smaller $\alpha$ will allow the rationality constraints to be overcome more easily by the data. One method to choose $\alpha$ is to use cross-validation.

## 5 Complexity and Constraint Decomposition

Clearly, the size of the rationality constraints is exponential in the number of variables. If there are $N$ players and $K$ pure strategies per player, there will be on the order of $NK$ variables, so the size of the constraint will be exponential in $NK$. Exponential cost in $N$ is acceptable, because the size of the normal form representation is itself exponential in the number of players. Furthermore, representations like graphical games provide a handle on how to deal with games with large numbers of players. However, an exponential cost in $K$ would severely limit the practicality of the approach. Fortunately, we can avoid this cost using constraint decomposition, which we now describe. We go through all four types of constraints in turn.

### 5.1 Strategy Maximum Likelihood Constraints

The strategy maximum likelihood constraints are unary constraints. In particular, there is one constraint for each pure strategy of each player. With $N$ players and $K$ pure strategies for each player, there are $NK$ strategy maximum likelihood constraints in our WCSP. If we use $\epsilon$ as the discretization parameter for strategy variables, then the complexity of the strategy maximum likelihood constraints is $\frac{NK}{\epsilon}$.

### 5.2 Strategy Consistency Constraints

Each strategy consistency constraint involves $K$ strategy variables, and there are $N$ such constraints. If we use $\epsilon$ as the discretization parameter for the strategy variables, then there are on the order of $\frac{1}{\epsilon}$ choices of values for each variable. For each player $i$, if we implement this consistency constraint directly, then there are $\frac{1}{\epsilon^K}$ values to search from where $\epsilon$ is a small positive real number. The complexity of the strategy consistency constraints is $\frac{N}{\epsilon^K}$ which is exponential in $K$.

However, for each constraint for each player $i$, we define $K$ extra variables $t_1, ..., t_k$ and decompose the multi-variable hard constraint into $K$ ternary hard constraints as follows.

$$\begin{cases} t_1 = \sigma_i(a_1) \\ t_2 = t_1 + \sigma_i(a_2) \\ ... \\ t_k = t_{k-1} + \sigma_i(a_n) = 1 \end{cases} \quad (10)$$

Given this decomposition, we only need to search from $\frac{K}{\epsilon^3}$ possible tuples of values. This is a significant reduction in complexity of this constraint as long as $K > 3$. The complexity of this constraint is $\frac{NK}{\epsilon^3}$ which is polynomial in $K$.

### 5.3 Payoff Maximum Likelihood Constraints

The payoff maximum likelihood constraints are unary constraints. There is one constraint for each pure strategy for each player $i$, for a total of $NK^N$ constraints. If we use $\delta$ as the discretization parameter for payoff variables, then the complexity of the payoff maximum likelihood constraint is $\frac{NK^N}{\delta}$.

### 5.4 Rationality Constraints

The rationality constraints are multi-variable soft constraints. There is one constraint for each pure strategy of each player $i$. With $N$ players and $K$ pure strategies for each player, we have $NK$ rationality constraints.

For each player $i$ and each pure strategy $a_k$ of player $i$, we can decompose the rationality constraints in several steps. First of all, the term $\sigma_{-i}(a_{-i}) u_i(a_k, a_{-i})$ can be decomposed into $(N+1)$ ternary constraints as follows:

$$\begin{cases} t_1 = \sigma_1(a_1) \\ ... \\ t_{i-1} = t_{i-2} \sigma_{i-1}(a_{i-1}) \\ t_{i+1} = t_{i-1} \sigma_{i+1}(a_{i+1}) \\ ... \\ t_n = t_{n-1} \sigma_n(a_n) \\ t_{n+1} = t_n u_i(a_k, a_{-i}) \end{cases} \quad (11)$$

Use $x_i$ to denote the last equation in the decomposition of the $i$th product $\sigma_{-i}(a_{-i})u_i(a_k, a_{-i})$. Then the term

$$\exp(\lambda \sum_{a_{-i}} \sigma_{-i}(a_{-i})u_i(a_k, a_{-i})) \quad (12)$$

can be decomposed as follows:

$$\begin{cases} y_1 = x_1 \\ y_2 = y_1 + x_2 \\ ... \\ y_{k^{n-1}} = y_{k^{n-1}-1} + x_{k^{n-1}} \\ y_{k^{n-1}+1} = e^{\lambda y_{k^{n-1}}} \end{cases} \quad (13)$$

For decomposing Equation 12, a set of constraints of in the form of equation 11 is constructed for every joint action profile $a_{-i}$. Since there are $K^{N-1}$ action profiles for $a_{-i}$, Equation 12 can be decomposed into $K^{N-1}(N+1)$ constraints. Notice that the decomposition of Equation 12 can be reused since it appears in two terms in the rationality constraint.

Finally, we decompose the following summation

$$\sum_{a_k \in A_i} \exp(\lambda \sum_{a_{-i}} \sigma_{-i}(a_{-i})u_i(a_k, a_{-i})) \quad (14)$$

This requires computing Equation 12 $K$ times for different values of $a_k$, plus another $K+1$ constraints for the sum. With one more constraint to express Equation 9, the total number of constraints is $K^N(N+1)+K+2$. Thus, the total number of constraints is on the order of $O(NK^N)$. If we use discretization parameter $\epsilon$ for strategy variables and $\delta$ for payoff variables, then the complexity of the rationality constraints is on the order of $O(\frac{NK^N}{\min\{\epsilon,\delta\}^3})$.

As a result of these decompositions, the size of the WCSP is polynomial in the size of the normal form game representation. Of course, WCSP is an NP-hard problem, so the complexity of our approach is probably exponential in the worst case. Nevertheless, WCSP is often tractable for real problems, so there is good reason to hope that it will be for us too.

# 6 Evaluation

In this section, we first discuss several baseline methods that we compare our learning method with. Then we describe our implementation, present our experimental results and discuss their implications.

## 6.1 Baseline Methods

A completely naive approach is to learn the strategies and payoffs of the game separately from the data given. We call this method the *Naive* method. This is a maximum likelihood approach for which the strategies and payoffs with highest log probabilities given the observations will be chosen. This approach respects the data completely. In our setting, the payoffs are noisily observed. Moreover, if the number of samples is small, then the observed frequency of each pure strategy might not reflect the underlying mixed strategy being played. For these reasons, we expect that the *Naive* method will perform poorly when the standard deviations of the payoffs are high and or when the number of samples is small.

A more sophisticated approach is to learn the payoff values from the observed data and infer the mixed strategies being played by solving for an equilibrium based on a certain solution concept. We use *NaiveNash* to refer to the method using the Nash equilibrium concept and *NaiveLQRE* to refer to the method using the LQRE concept. This approach ignores the observed pure strategy profiles.

## 6.2 Experimental Setup

We evaluate our learning methods with computational experiments on randomly generated normal form games with 2 players and 2 pure strategies for each player (referred to as 2-player game hereafter). Our implementation consists of the following steps:
1. Generate a random 2-player normal form game.
2. Use Gambit (McKelvey et al., 2007) to compute LQRE of the game with different $\lambda$ values.
3. Given specific LQRE strategies, generate data on game play, assuming players are playing according to the specified strategies, and payoffs are generated from a Gaussian distribution centered around the payoffs specified in the game.
4. Generate and store a WCSP problem in a predefined XML format. This XML file defines the variables, the domains of the variables, and the constraints in the form of costs associated with value assignment tuples.
5. Solve the WCSP specified in the XML file using Toolbar2 (Bouveret et al., 2004) which is a state-of-the-art C++ solver for WCSP, Max-SAT, and Bayesian Networks.

We derived our results by averaging the data over 10 random 2-player normal form games. The payoff values are sampled uniformly from $[1,2]$. We chose the strategy discretization parameter $\epsilon$ to be 5%, the payoff discretization parameter $\delta$ to be 0.1, the LQRE multiplier $\alpha$ to be 100, and the standard deviation $R$ of payoff values to be 0.7. These are parameters of our model that need to be set manually and finding ways to set them automatically will be future work. For our preliminary results, the implementation does not include the constraint decomposition described. We had to truncate the LQRE constraint written out into the XML file to only contain $\frac{1}{5}$ of all the generated tuples with the smallest costs due to the enormous size of the resulting XML file (70mb). Although this does not affect the validity of our results, we believe that incorporating the constraint decomposition will result in an immediate improvement for the efficiency of our implementation.

For the *Naive* method, the XML file includes only the strategy and payoff maximum likelihood constraints. For the NaiveNash method, we take the payoff values learned from the *Naive* method, use Gambit to solve for Nash equilibrium. For games with multiple Nash equilibria, we pick the one minimizing our error measure. For the NaiveLQRE method, we again take the payoff values learned from the *Naive* method and use Gambit to solve for the LQRE using the same $\lambda$ value used to generate the data.

### 6.3 Results and Discussion

We evaluated our learning method in three different settings. First, we varied the training set sizes and compare our approach with the *Naive*, *NaiveLQRE*, and *NaiveNash* methods. Second, we varied the $\lambda$ values and do the same comparison. Finally, we examined the performance of our algorithm using wrong $\lambda$. The error of each learning instance is measured by computing the Euclidean distance between the combined vectors of actual versus learned strategy and payoff values.

Varying training set sizes

We first consider varying the training set size, i.e. the number of samples of the game plays observed. When the training set size is small, learning an accurate representation of the game is expected to be difficult since the limited amount of data does not reflect the mixed strategy chosen by the players accurately. Also, with limited amount of data, the payoff values sampled with noise may be too biased to reflect the true mean and standard deviation of the observed payoff values.

In this setting, we found that our approach with rationality constraints performs better than the naive approach when the training set size is small. Table 1 summarizes our results for this setting. When the number of samples is 10, the average error obtained by our approach is 8% less than that of the *Naive* method. Also, we found that the *NaiveLQRE* and *NaiveNash* perform considerably worse than the *Naive* method and our LQRE method. This observation is expected, especially for NaiveNash, since the limited amount of noisy data causes the maximum likelihood methods to produce inaccurate payoff values. As a result, solving for equilibrium strategies using these payoff values result in equilibrium strategies that tend to be far from the actual strategies being played.

However, as the number of samples increases, we observe that the *Naive* method gradually outperforms our LQRE approach since there is enough data to for the *Naive* method to learn the payoff values accurately. By learning accurate payoff values, the errors of *NaiveLQRE* and the *NaiveNash* methods also improve. In particular, when the number of samples is 50, the average error of our LQRE approach is still close to that of the *Naive* method. However, for 100 samples, the *Naive* method outperforms our LQRE approach by a larger amount.

Table 1: Errors for Varying Training Set Sizes, $\lambda \approx 3$

|           | M = 10 | M = 50 | M = 100 |
|-----------|--------|--------|---------|
| LQRE      | 0.994  | 0.684  | 0.642   |
| Naive     | 1.086  | 0.668  | 0.473   |
| NaiveLQRE | 1.117  | 0.729  | 0.516   |
| NaiveNash | 1.266  | 0.990  | 0.812   |

Varying $\lambda$ values

Next, we tested how well our LQRE approach performs as we vary the value of $\lambda$. For this set of results, we fix the number of samples to be 10 and vary the $\lambda$ values used to solve for the LQRE of the games. The concern of this setting is whether the correct $\lambda$ value will guide the WCSP with rationality constraints to find the correct mixed strategy and payoff combinations.

We ran experiments for 3 $\lambda$ values and the results are shown in Table 2. The lambda values are approximate since Gambit does not allow us to specify the lambda values with which to solve the LQRE of the game. Our findings indicate that our approach performs better than the naive approach for small and moderate values of $\lambda$, but not for very large $\lambda$. We believe that this is because the LQRE approaches Nash equilibrium as $\lambda$ approaches infinity. For the games we considered, we observed that when $\lambda$ is around 10, the mixed strategies corresponding to a LQRE are already extremely close to the Nash equilibrium strategies of the game.

Table 2: Errors for Varying $\lambda$ Values, $M = 10$

|           | $\lambda \approx 1$ | $\lambda \approx 3$ | $\lambda \approx 10$ |
|-----------|---------------------|---------------------|----------------------|
| LQRE      | 0.954               | 0.994               | 1.202                |
| Naive     | 1.069               | 1.086               | 1.078                |
| NaiveLQRE | 1.037               | 1.117               | 1.130                |
| NaiveNash | 1.325               | 1.266               | 1.184                |

Using the wrong $\lambda$

A potential criticism of our approach is that it requires the knowledge of $\lambda$ or how approximately rational the players are. It might be unrealistic to expect this in real world examples. To counter this criticism, we investigated how well our approach works when we have the wrong $\lambda$ value, i.e., when the $\lambda$ used for generating the data is different from the $\lambda$ used for learning. In other words, does it help to assume that the agents are somewhat rational, even if we are wrong about to the degree of their rationality? To test this question, we ran experiments fixing $M = 10$ and $\lambda = 1$ for learning but varying the actual $\lambda$ used to generate data.

The results are shown in Table 3. The column headers are the actual values of $\lambda$ used to generate the data. We were pleasantly surprised to find that our algorithm performs quite well even when $\lambda$ is significantly wrong, and the performance does not degrade quickly. When the true $\lambda$ is 1, which means that we are learning with the correct value, our method improves over the naive method by 13%. When the true $\lambda$ is 2, our method still improves over the naive method by 9%. This indicates that imposing rationality constraints is beneficial, even when it is not exactly correct.

Table 3: Errors for Using Wrong $\lambda$ Values, $M = 10$

|  | $\lambda \approx 0.5$ | $\lambda \approx 1$ | $\lambda \approx 1.5$ | $\lambda \approx 2$ |
|---|---|---|---|---|
| LQRE | 0.856 | 0.877 | 0.927 | 0.916 |
| Naive | 0.986 | 1.010 | 1.036 | 1.008 |
| % Improvement | 13.18 | 13.17 | 10.52 | 9.13 |

## 7 Conclusion and Future Work

This paper has introduced the idea that connecting strategies to payoffs using rationality constraints can improve learning games from data, particularly with limited data. While we have only demonstrated this for 2-by-2 normal form games, the paper serves as a proof by demonstration that the idea can be beneficial. Although our experimental results are preliminary, we hope that this paper opens the door to research on efficient and scalable implementations and approximations of the idea. For future work, we first would like to generate data for games with more players, more actions for each player, and greater number of samples. We also plan to experiment with games with interesting structures. There are several other specific future directions. It would be preferable to make $\lambda$ a variable that is learned rather than treating it as a fixed parameter that must be set. Making $\lambda$ a variable can be accommodated in our WCSP framework. Also, we might not want to assume that all agents use the same $\lambda$, and our LQRE formulation can be generalized to capture this. Moreover, in addition to using wrong $\lambda$ values, it would be worthwhile to investigate the case when our LQRE assumption is incorrect. Alternatively, a possible extension is to combine the rationality constraints, which as we have argued are like Bayesian priors, with Bayesian priors of the traditional kind.

## Acknowledgements

This work was supported by the Air Force Office of Scientific Research under MURI contract 5710002613.